\documentclass[conference]{IEEEtran}
\IEEEoverridecommandlockouts
\usepackage{cite}
\usepackage{amsmath,amssymb,amsfonts}
\usepackage{tikz}
\usepackage{algorithmic}
\usepackage[utf8]{inputenc}
\usepackage[T1]{fontenc} 
\usepackage{graphicx}
\usepackage{textcomp}
\usepackage{hyperref}
\usepackage{dblfloatfix}
\usepackage{tikz}
	\usetikzlibrary{shapes.geometric}
\usepackage{xcolor}
\usepackage{array}
\hypersetup{
	hidelinks, 
}

\begin{document}
\title{DeFi Risk Transfer: \\ Towards A Fully Decentralized Insurance Protocol}

\author{
\IEEEauthorblockN{Matthias Nadler}
\IEEEauthorblockA{\textit{Center for Innovative Finance (CIF)} \\
\textit{University of Basel}\\
Basel, Switzerland \\
matthias.nadler@unibas.ch}
\and
\IEEEauthorblockN{Felix Bekemeier}
\IEEEauthorblockA{\textit{Center for Innovative Finance (CIF)} \\
\textit{University of Basel}\\
Basel, Switzerland \\
felix.bekemeier@unibas.ch}
\and
\IEEEauthorblockN{Fabian Schär}
\IEEEauthorblockA{\textit{Center for Innovative Finance (CIF)} \\
\textit{University of Basel}\\
Basel, Switzerland \\
f.schaer@unibas.ch}
}

\maketitle
\begin{abstract}
In this paper, we propose a fully decentralized and smart contract-based insurance protocol. We identify various issues in the Decentralized Finance (DeFi) insurance context and propose a solution to overcome these shortcomings. We introduce an economic model that allows for risk transfer without any external dependencies or centralized intermediaries. In particular, our proposal does not need any sort of subjective claim assessment, community voting or external data providers (oracles). 
Moreover, it solves the problem of over-insurance and proposes various ways to mitigate the capital inefficiencies usually seen with DeFi collateral. 
The work takes inspiration from peer-to-peer (P2P) insurance and collateralized debt obligations (CDO). 
We formally describe the protocol, assess its efficiency and key properties and present a reference implementation.
Finally, we address limitations, extensions and ideas for further research. 
\end{abstract}


\section{Introduction}
\label{sec:introduction}

Decentralized Finance (DeFi) refers to public blockchain-based financial infrastructure that uses smart contracts to replicate traditional financial services in a more open, interoperable, and transparent way \cite{Schaer.2021}. These smart contract-based services are usually referred to as protocols. They provide basic building blocks such as the opportunity to swap assets or allocate liquidity efficiently and can be reused and combined in any way. While decentralized exchanges and lending markets are arguably among the most prominent protocols and get a lot of attention, there are other crucial building blocks that are required for a well-functioning financial infrastructure. One of these building blocks is the ability to transfer risks.  

Consider the following general example: An economic agent has an investment opportunity that may result in a small loss or a large gain. Further assume that both outcomes have the same probability. The expected return would be positive and a risk-neutral (or risk-seeking) agent would be willing to engage. However, if the same opportunity is instead presented to a risk-averse agent, they may decline and forego a positive expected return due to the cost of uncertainty. If a financial market allows risk to be transferred, there is a simple solution. The risk-averse person can approach an entity with a higher risk tolerance and offer them a premium in return for their willingness to bear the risk. They essentially share the positive expected return and the risk would be borne by the entity with the higher risk tolerance. 

Similarly, a blockchain-based financial infrastructure becomes more efficient if smart contract risks are transferable. Risk-averse investors could share some of their expected return as a compensation for an insurance policy that covers the smart contract risks of the respective liquidity pool. DeFi users who are willing to bear additional risk could generate a higher yield. The existence of a market for risk transfer would be beneficial for everyone, as it allows all DeFi users to structure their portfolio in accordance with their individual risk preferences. 

There already exists a relatively large number of smart contract-based insurance protocols, including but not limited to Nexus Mutual \cite{Karp.2017}, Nsure \cite{Nsure.2020}, cozy.finance \cite{CozyFinance.2020}, Unslashed Finance \cite{Unslashed.2021} and Risk Harbor \cite{riskHarbor2}. While some of these protocols offer innovative solutions and have provided valuable contributions to the DeFi protocol space, they are arguably not fully decentralized and face various challenges.

\textit{First}, insurance requires that the insurer can credibly demonstrate its ability to cover potential losses at all times. Centralized insurance is based on a combination of reputation and regulation. Moreover, centralized insurance companies rely on active asset and risk management to strike a balance between liquidity and capital efficiency. DeFi, on the other hand, is built on a pseudonymous system with little to no legal recourse. It relies on transparency and (over-)collateralization. Consequently, many implementations face trade-offs between capital efficiency, security and special privileges that allow for manual interventions.

\textit{Second}, DeFi insurance protocols usually struggle with claim assessment. Generally speaking there are two options. (a) The insurance policy is parametric and relies on oracles and (b) the outcome is decided through a vote, by so-called claim assessors. 
Both approaches are quite subjective and can easily lead to false outcomes. The former introduces dependencies to external data providers and does not reflect true damages due to its parametric nature. The latter relies on a voting process among pseudonymous actors that can assume various roles within (and outside) the system. Moreover, truly decentralized voting will be either subject to sybil attacks \cite{douceur2002sybil} or whale dominance with potentially problematic incentives. 
There are good arguments, why neither the oracle-based nor the claim assessor-based approach should be considered fully decentralized.

\textit{Third}, most protocols cannot prevent over-insurance. DeFi users can buy cover for protocols to which they have no exposure. This can create problematic incentives and -- depending on the jurisdiction -- result in conflict with the law.

In this paper, we propose a novel DeFi insurance protocol that solves these issues. To the best of our knowledge it is the first proposal for a fully decentralized insurance protocol with no external dependencies. 
As part of this research project, we have also built a basic reference implementation of the protocol. The implementation can be found in the appendix.

After this short introduction, we discuss related works from the DeFi, insurance and finance literature. In Section \ref{sec:protocol} we turn to the technical part, describe the protocol and perform a gas efficiency analysis. In Section \ref{sec:divergence} we study external incentives for liquidity providers and derive the implicit cost of liquidity provision for various pools involving our protocol's tranche tokens. In Section \ref{sec:discussion} we discuss our results, potential extensions and limitations. Finally, we conclude in Section \ref{sec:conclusion}.

\section{Related work}
\label{sec:literature}

The motivation for a DeFi insurance protocol is closely linked to discussions on smart contract and DeFi risks, protocol failures and shock propagation. These issues have received an increasing amount of research attention and are an important part of the academic discourse on DeFi \cite{Atzei.2017,Gudgeon.2020,Macrinici.2018,Zheng.2020,Zhou.2022}. 
Our protocol can mitigate some of the consequences by allocating risk in a more efficient way. Moreover, market prices for risk premiums can serve as an indication of the perceived risk; similar to prediction markets.
With regard to yield-generating lending protocols, different authors discuss the risks of illiquidity, dependencies and misaligned incentives \cite{Gudgeon.2020, Bartoletti.2021, Lehar.2022, Qin.2021}. Moreover, there are various papers discussing oracle reliability and potential manipulation \cite{Angeris.2020, Liu.2021}. 
Our proposal does not have any dependencies, allows the insurant to hedge against oracle exposure, and even works in situations where the insured protocols become illiquid.

Existing DeFi insurance protocols are mostly based on principles of mutual insurance, where users participate in the commercial success of the protocol. In theory, mutuals can have certain advantages for large risk pools \cite{Albrecht.2017}, in the presence of transactional costs and governance issues \cite{Laux.2010}, and in addressing problems of adverse selection \cite{Ligon.2005}. However, due to centralized economic value capture in most mutuals, problems potentially remain with respect to default risks \cite{Tapiero.1986}. 
In a DeFi context, mutual-based insurance protocols usually rely on centralized or vote-based claim assessment and may depend on \emph{know your customer} (KYC) principles or introduce other forms of dependencies.

Our protocol is fundamentally different from a mutual insurance. There is no centralized economic value capture and the protocol does not accumulate reserves. The general concept of our protocol is inspired by peer-to-peer (P2P) insurance and financial instruments with tranches, such as collateralized debt obligations (CDO). 

In a P2P insurance model, individuals pool their insurance premiums and use these funds to cover individual damages. P2P risk transfer is still at a very early stage of research, with seminal works including \cite{Denuit.2022,Denuit.2020b,Feng.2022b,Feng.2022,Denuit.2019, Denuit.2020}. Several authors have started to formally explore the organizational structure, optimality and pitfalls of P2P insurance \cite{Charpentier.2021,Clemente.2020,Levantesi.2022}.

Our protocol is based on similar principles. In particular, we make use of different risk preferences and levels that allow individuals to pool their risks without the explicit need for an intermediary. 
However, there is an important difference between P2P insurance and our approach: P2P insurance usually covers individual risks. As such, P2P insurance is built on the general assumption that damages within the collective are uncorrelated and that premiums of the unaffected insurants can be used to compensate the ones that have suffered losses. Our protocol insures large scale risks that will affect all insurance holders. Consequently, we need explicit roles in accordance with the individuals' risk preferences. This is achieved by creating tranches with different seniorities and security guarantees.

As such, our protocol incorporates some aspects of CDOs. CDOs have been discussed extensively in the subject-related literature \cite{duffie2001risk, armstrong2005understanding, lucas2006collateralized, bluhm2011valuation}. They split cash flows among tranches with different seniority. The most senior tranches are honored first and the most junior tranches bear the losses.
In addition to traditional use cases, such as CDOs for bank refinancing, insurance risk also appears to be a suitable use-case for CDOs \cite{forrester2008insurance}. 
Likewise, CDOs are used widely in various applications, also outside traditional financial markets. For example, CDOs have already been discussed related to the support of microcredits \cite{bystroem2008microfinance}.

This combination of P2P insurance, seniority-based promises and DeFi specifics builds the foundation of our protocol and allow us to propose a fully decentralized DeFi insurance.

\section{Protocol}
\label{sec:protocol}

In this section, we present a decentralized risk hedging protocol, based on tranched insurance. First, we provide a quick overview and describe the core functionality of the protocol. Second, we take a more technical perspective and describe individual function calls and state transitions. Third, we discuss potential technical extensions and trade-offs. Fourth, we provide a short efficiency analysis and discussion of the protocol's computational costs (gas fees).

\subsection{Protocol Overview}
\label{sub:overview}
The general idea of our insurance protocol is to pool assets from two third-party protocols, and allow users to split the pool redemption rights into two tranches: $A$ and $B$. If any of the third-party protocols suffer losses during the insurance period, those losses will be primarily borne by the $B$-tranche holders. $A$-tranche holders will only be negatively affected if 50\% or more of the pooled funds are irrecoverable, or if both protocols become temporarily illiquid and face (partial) losses. 
We effectively split the redemption rights into a riskier and less risky version and allow the market for $A$- and $B$-tranches to determine the fair risk premium in line with the users' expectations.

The protocol consists of three main phases: \textit{risk splitting}, \textit{investing/divesting} and \textit{redemption}.

In the \textbf{risk splitting} phase, anyone may allocate their preferred number of $C$-tokens to the insurance protocol. These $C$-tokens represent the underlying asset, e.g., a stablecoin or Ether. In exchange, the users receive equal denominations in $A$- and $B$-tranches, thereby ensuring that an equal number of both tranches will be created. $A$ and $B$ are ERC-20 compliant tokens and can be transferred separately. This allows the users to swap the tokens on decentralized exchanges to obtain a relative allocation of $A$- and $B$-tranches that reflects their risk preferences.

At the beginning of the \textbf{invest/divest} phase, the insurance protocol allocates the accumulated collateral of $C$ equally into two protocols. 
In return, it receives interest-bearing tokens (wrapped liquidity shares) from each protocol. We denote these shares as $C_x$ and $C_y$. 

To make things less abstract, consider the following example: A stablecoin ($C$) gets allocated to two distinct yield-generating lending protocols. In return, the insurance protocol receives the respective interest-bearing tokens ($C_x$ and $C_y$). They are locked in the insurance contract, where they will accumulate interest over time.
At the end of the invest/divest phase, the insurance protocol tries to liquidate the wrapped shares. This is a necessary step in preparation for the redemption of the $A$- and $B$-tranches.  

In a third step, the protocol enters the \textbf{redemption} phase. The goal of this phase is to compute potential losses and allow the $A$- and $B$-tranche holders to claim their respective share of the underlying. 
It is important to understand that the redemption phase can be executed in one of two distinct modes. Mode selection depends on the success of the liquidation at the end of the invest/divest phase.
If the liquidation of $C_x$ and $C_y$ works as expected and the insurance protocol receives the collateral tokens $C$, then redemption can be conducted in \textit{liquid mode}. In this mode, it is straightforward to distribute the interest equally among all $A$- and $B$-tranche holders. Similarly, potential losses can be computed and primarily allocated to $B$-tranche holders. 
If the liquidation of $C_x$ or $C_y$ fails, the protocol enters \textit{fallback mode}. This can happen if a third-party protocol suffers from a liquidity crunch or if an external contract changes the expected behavior. 
In fallback mode, users redeem their tranche tokens directly for their preferred mix of $C_x$ and $C_y$ tokens. The higher tranche seniority of $A$-tranches is ensured through a timelock-based redemption sequence. In a first step, $A$-tranche holders get to choose if they want to claim their share in $C_x$, $C_y$ or a mix of the two. After the timelock is over, $B$-tranche holders can claim what is left.

\subsection{Technical Implementation}
\label{sub:technical}

\begin{figure*}
	\begin{tikzpicture}[align=center, node distance=5cm, scale=0.6, every node/.style={scale=0.6}]
		
		\tikzstyle{arrow} = [thick, ->, >=stealth]
		\tikzstyle{box} = [rectangle, thick, rounded corners, draw = black, minimum height = 1.5cm, minimum width = 3cm]
		
		\coordinate (0) at (0, 0);
		\draw[fill=black] (0) circle (5pt);
		
		\node [box] (1) [right of=0, xshift = -1cm] {\textbf{ReadyToAccept} \\ $\bullet$ \texttt{splitRisk()}};
		\node [box] (2) [right of=1, xshift = 1cm]  {\textbf{ReadyToInvest} \\ $\bullet$ {\texttt{invest()}}};
		\node [box] (3) [right of=2, xshift = 1cm] {\textbf{MainCoverActive}};
		\node [box] (4) [right of=3, xshift = 1cm] {\textbf{ReadyToDivest} \\ $\bullet$ {\texttt{divest()}}};
		\node [box] (6) [right of=4, xshift = 1cm] {\textbf{FallbackOnlyA}\\ $\bullet$ \texttt{claimA()}};
		\node [box] (5) [above of=6, yshift=-2.5cm] {\textbf{Liquid} \\ $\bullet$ \texttt{claimAll()} \\ $\bullet$ \texttt{claim()}};
		\node [box] (7) [below of=6, yshift = 2.5cm] {\textbf{FallbackAll}\\ $\bullet$ \texttt{claimA()} \\ $\bullet$ \texttt{claimB()}};
		
		\coordinate (X) at (10, 2.8);
		\coordinate (Y) at (26.5, 2.2);
		\coordinate (Z) at (26.5, 2.8);
		
		\draw [arrow] (0) -- node [above] {Deployment} 	(1);
		\draw [arrow] (1) -- node [above, align=center] {\small{\texttt{block.timestamp}}\\ $=S$} (2);
		\draw [arrow] (2) -- node [above, align=center] {\small{\texttt{invest()}}\\successful}		(3);
		\draw [arrow] (3) -- node [above, align=center] {\small{\texttt{block.timestamp}}\\  $=T_1$}	(4);
		\draw [arrow] (22,0.75) -- (22,2.2) -- node [below, align=left, yshift =0cm] {\small{\texttt{divest()}}\\ successful}	(Y);
		\draw [arrow] (6) -- node [left, align=center] {\small{\texttt{block.timestamp}}\\  $=T_3$}	(7);
		\draw [arrow] (4) -- node [above, align=center, yshift=0cm] {\small{\texttt{block.timestamp}}\\ $=T_2$} (6);
		

		
		\draw[thick, arrow] (2) -- node [right, yshift=0.4cm] {\small{\texttt{block.timestamp}}\\  $=T_1$} (X) -- (Z);
		
	\end{tikzpicture}
	
	\caption{State Transition Diagram: Represents state transitions and their respective function sets.}
	\label{fig:states}
\end{figure*}
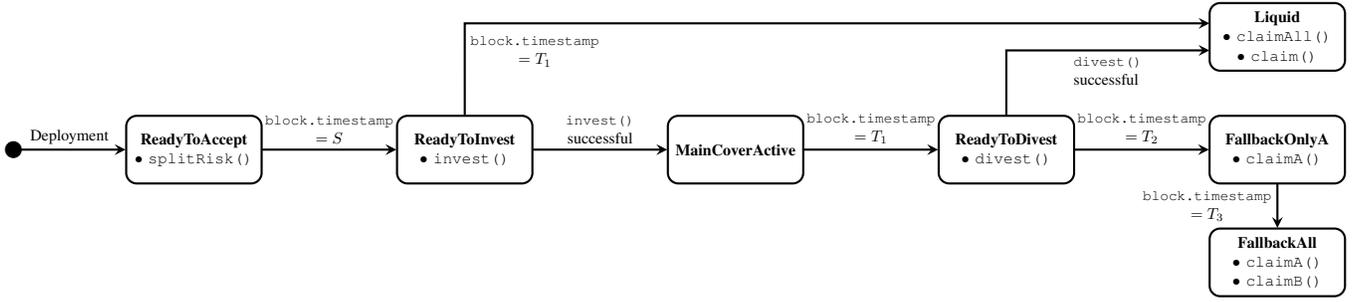

A reference implementation of the insurance contract is available in the appendix and demonstrates how our protocol can be used to provide insurance for two yield-generating protocols that wrap the Maker DAO stablecoin Dai \cite{Maker.2017}, denoted as $C$. The two yield-generating protocols are Aave version 2 \cite{Aave.2020} with aDai and Compound Finance \cite{Compound.2019} with cDai, denoted as $C_x$ and $C_y$ respectively. The reference implementation includes the full Solidity code for the Ethereum Virtual Machine-based (EVM) contract and can be used as a starting point for developers who want to create their own insurance contracts using a similar approach.

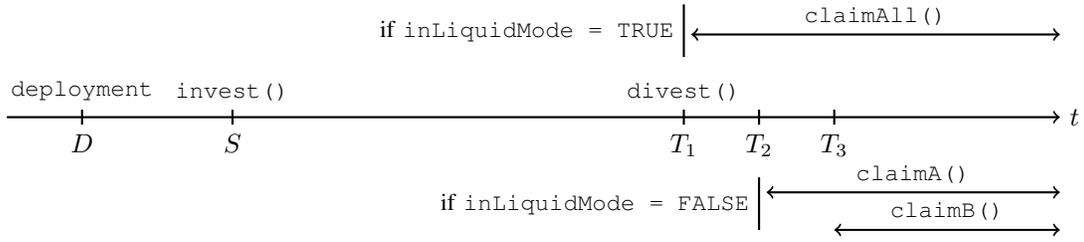
\begin{figure*}[h]
\centering
\begin{tikzpicture}[thick]
    \draw [->,thick] (0,0) -- (14,0) node (time) [right] {$t$};
    
    \draw (1,-0.1) node [below, align=center] {$D$} node [above=0.2cm] {\small{\texttt{deployment}}}-- (1,0.1) ;
    \draw (3,-0.1) node [below] {$S$} node [above=0.2cm] {\small{\texttt{invest()}}}-- (3,0.1) ;
    \draw (9,-0.1) node [below] {$T_{1}$} node [above=0.2cm] {\small{\texttt{divest()}}}-- (9,0.1) ;

    \draw (9,0.8) -- (9,1.5) node [left, midway] {\small{if \texttt{inLiquidMode = TRUE}}};    
    \draw[<->] (9.1,1.1)    -- (14,1.1)node [midway, above] {\small{\texttt{claimAll()}}} ;
    
    \draw (10,-0.8) -- (10,-1.5) node [left, midway] {\small{if \texttt{inLiquidMode = FALSE}}};    
    \draw (10,-0.1) node [below] {$T_{2}$} -- (10,0.1) ;
    \draw (11,-0.1) node [below] {$T_{3}$} -- (11,0.1) ;
    
    \draw[<->] (10.1,-1)    -- (14,-1)node [midway, above] {\small{\texttt{claimA()}}} ;
    \draw[<->] (11,-1.5)  -- (14,-1.5)node [midway, above] {\small{\texttt{claimB()}}} ;
    
\end{tikzpicture}
\caption{Sequential actions in liquid mode (top, \texttt{divest()} successful) and fallback mode (bottom, \texttt{divest()} unsuccessful).}
\label{fig:sequential}
\end{figure*}

In this subsection we provide an overview of the reference implementation's technical specifications, including the functions, variables and states. We present this information in a chronological order, following the timeline presented in Figure \ref{fig:sequential}.
The states are referred to as: \textit{ReadyToAccept, ReadyToInvest, MainCoverActive, ReadyToDivest, Liquid, FallbackOnlyA} and \textit{FallbackAll}. 
Note that strictly speaking a smart contract cannot automatically transition from one state to another based on the passage of time; this is a fundamental limitation of smart contract technology. Any state change on the contract has to be initiated by a function call. Our implementation works around this by defining states as a set of successfully callable functions and reverting function calls, if they are outside the allowed time windows. Hence, the set may change based on time conditions.

Before the first state, the initial parameters must be defined and contract deployed. The parameters include the addresses of the tokens involved in the contract, as well as the absolute values for the timestamps when state transitions occur. These forced state transitions are represented in Figures \ref{fig:states} and \ref{fig:sequential} as $S$, $T_1$, $T_2$ and $T_3$, where  $S < T_1 < T_2 < T_3$. 
Furthermore, the constructor deploys two ERC-20 token contracts for $A$- and $B$-tranches, with the insurance contract as the sole, immutable owner. This means that only the insurance contract can mint and burn the tranche tokens.

After deployment, the contract is in the \texttt{ReadyToAccept} state and the public function \texttt{splitRisk()} is available for anyone to call. The input parameter for the function is an amount of $C$ tokens. The \texttt{splitRisk()} function then transfers this amount of $C$ tokens from the caller to the insurance contract and issues a number of $A$- and $B$-tranche tokens equal to half that amount to the caller. For example, if the input is 100, the function will transfer 100 $C$ tokens from the caller to the insurance contract and issue 50 tranche $A$ tokens and 50 tranche $B$ tokens to the caller. It is important to note that the act of calling the \texttt{splitRisk()} function does not provide the user with any form of insurance cover. In order to obtain insurance cover -- or to assume more risk -- the user must sell or trade a portion of their tranche $A$ or tranche $B$ tokens.

When time $S$ is reached, the contract transitions to the \texttt{ReadyToInvest} state and users can no longer mint new tranche tokens. The \texttt{invest()} function is available during this state and it is tailored to the specific needs of the protocols that are part of the insurance contract, with the goal of splitting the deposited $C$ tokens equally among the protocols. In the reference implementation, the function will send half of the available $C$ to Aave and the other half to Compound in exchange for their respective yield-bearing tokens, $C_x$ and $C_y$. After a successful \texttt{invest()} call, the insurance contact holds $C_x$ and $C_y$ of equal value and does no longer hold $C$. Calling the invest function incurs a transaction fee, paid by the caller while the benefits of the call are shared among all participants. To avoid the problem of a first mover disadvantage, to ensure that the call is executed in a timely fashion and to split the costs equally among all participants, the \texttt{invest()} function should compensate the caller for executing the transaction.\footnote{We did not include a compensation mechanism in the reference implementation. When implemented, it should cover at least the base fee of the transaction plus a fixed amount for the tip.} The unlikely case in which no successful \texttt{invest()} call is made before the forced state transition at $T_1$ will be covered later in this subsection. 

When a successful \texttt{invest()} call is made, the contract transitions to the \texttt{MainCoverActive} state and sets the variable \texttt{isInvested = true}. The contract is now exposed to the risks of the third-party protocols and the main period of insurance cover for the $A$-tranches begins. In this state, no functions can be called on the contract. However, the $A$- and $B$-tranches remain transferable.

At time $T_1$, the contract will transition from the \texttt{MainCoverActive} state to the \texttt{ReadyToDivest} state, where the \texttt{divest()} function can be invoked. It has a similar structure to the \texttt{invest()} function, but instead of depositing the underlying assets into the third-party protocols, \texttt{divest()} tries to withdraw the underlying assets including any accumulated yield from the protocols. A \texttt{divest()} call is considered successful if no errors occur while withdrawing the assets and if both $C_x$ and $C_y$ have been fully converted back to $C$.

\begin{table*}[b!] 
\center
\caption{The three potential outcomes for liquid mode}
\label{tbl:liquid}
\setlength{\extrarowheight}{5pt}
\begin{tabular}{c c c p{8cm}}
\hline \hline
	\textbf{Case} & \textbf{Payoff $\mathbf{A}$} &  \textbf{Payoff $\mathbf{B}$} & \textbf{Description} \\ \hline 
	$C_{T_1} \geq C_S$ & $ \frac{C_{T_1}}{2}$ & $ \frac{C_{T_1}}{2}$ & Proceeds are split equally among all tranche token holders. Both tranches are treated equally.\\\hline
	$ C_S > C_{T_1} > \frac{C_S}{2}$ & $ \frac{C_S}{2} + i$ & $C_{T_{1}} - \left( \frac{C_{S}}{2} + i \right) $  & $A$-tranche holders get fully compensated and receive yield payment. $B$-tranche holders receive a proportion of their initial stake.\\ \hline
	$C_{T_1} \leq  \frac{C_S}{2}$ & $C_{T_1}$ & $0$ & Proceeds are used to partially compensate $A$-tranche holders. This can only occur if both yield-generating protocols suffer losses.   \\ \hline \hline
\end{tabular} 
\end{table*}

A successful \texttt{divest()} call immediately transitions the protocol to the \texttt{Liquid} state by setting \texttt{inLiquidMode = true}. In this state, the allocation of the redeemed assets to the $A$- and $B$-tranches is deterministic and can be calculated as part of the \texttt{divest()} call. Let us define $C_S$ as the total initially invested amount, $C_{T_1}$ as the total redeemed amount and $i$ as the interest. We can then differentiate between three cases and determine the payouts for each case, as shown in Table \ref{tbl:liquid}. The payout per $A$- and $B$-tranche token is stored on the contract and can be accessed using the variables $cPayoutA$ and $cPayoutB$, respectively. During the liquid state, users can call the \texttt{claim()} function which accepts an amount for $A$- and $B$-tranches as input. If the caller is in control of at least the specified amount of tranches, the contract will burn these tranches and transfer the payout to the caller. For convenience, a \texttt{claimAll()} function is available and will internally call the \texttt{claim()} function with the caller's current balance of tranches.

If no successful \texttt{divest()} call is made during the \texttt{ReadyToDivest} state, a forced transition occurs at $T_2$ and the protocol enters fallback mode, which starts in state \texttt{FallbackOnlyA}. In fallback mode, the protocol has no knowledge about the value of its interest-bearing tokens relative to the initial investment. Therefore, instead of assigning a payout to the tranches, the tranche holders can choose which of the two interest-bearing tokens they would like to redeem.

Based on the total amount of tranche tokens and the remaining interest-bearing tokens, the contract determines a fixed redeem-ratio for each of the two interest-bearing tokens. These ratios are stored on the contract as $cxPayout$ and $cyPayout$ and are defined as the total amount of the respective asset, divided by half of the total amount of tranches. For example, assume $50$ $A$- and $50$ $B$-token have been minted and the contract holds $20$ $C_x$ and $1500$ $C_y$. A tranche can now be redeemed for $0.4$ $C_x$ or $30$ $C_y$. Once all tranches are redeemed there are no interest-bearing tokens left on the contract.
$A$- and $B$-tranches can be redeemed for the same amount. However, during the \texttt{FallbackOnlyA} state, as the name suggests, only $A$-tranches can be redeemed for interest-bearing tokens with the function \texttt{claimA()}. As an input for this function, the caller specifies how many of their $A$-tranches they want to redeem for $C_x$ and how many for $C_y$. The contract then burns the tranches and transfers the assets according to the redeem-ratios.
 
At time $T_3$, if the contract is in fallback mode, the final transition happens to the \texttt{FallbackAll} state. This state is identical to \texttt{FallbackOnlyA} with the only difference that $B$-tranches can now also be redeemed via the \texttt{claimB()} function.
 
Finally, to ensure we never end up in a state where the assets cannot be recovered, we need to define a state transition from \texttt{ReadyToInvest} to \texttt{Liquid} if the \texttt{invest()} function was not successfully called. This transition happens after $T_1$ if \texttt{isInvested == false} and allows the users to reclaim their initially invested funds.

\subsection{Extensions and Trade-Offs}
\label{sub:extensions}
To obtain insurance cover, a protocol user must sell their $B$-tranches. A possible extension to the insurance contract would be to use intra-transaction composability and connect it to a decentralized exchange. This would allow users to sell their $B$-tranches in the same transaction as the \texttt{splitRisk()} function. However, note that any additions to the insurance contract will introduce additional risk. Keeping the contract as simple as possible and reducing dependencies to a minimum will help to manage this risk. We argue that most extensions which introduce new dependencies should be implemented at the user interface level in a separate contract.
 
Consider the following example: Let us assume that we want to create a function to insure an amount of $C$ tokens. We create a new contract with a function that uses a flash loan \cite{Qin.2021} for twice the amount and calls \texttt{splitRisk()}. In the same function, the $B$-tranches are sold to a decentralized exchange and the $A$-tranches transferred to the caller. Finally, the flash loan is repaid, using the proceeds from the sale and the funds from the initial caller.
The additional contract can be developed and deployed independently of the insurance contract. This separation offers more flexibility and introduces no additional risks for other users. The trade-off here is that the transaction fees might be slightly higher, as external calls are more costly than internal ones.

\subsection{Transaction Costs}



Depositing funds into a protocol incurs a transaction fee, which is imposed by the blockchain network and expressed in units of computation -- commonly called gas. This transaction fee can vary slightly based on circumstantial parameters, but it largely depends on the computational complexity of the transaction. Depositing funds into our reference implementation via the \texttt{splitRisk()} function costs around 83,000 gas. Depositing to Aave or Compound directly incurs a fee of 249,000 or 156,000 gas, respectively. While calling the \texttt{invest()} function is expensive (488,000 gas), this cost can be split among all users in the insurance contract. Similar to yield aggregation protocols \cite{Cousaert.2022}, the insurance contract becomes more gas efficient, the more users participate and even for just a few users, we expect the minting of insured tokens to be cheaper than minting uninsured tokens.

\section{LP-Incentives and Divergence Loss}
\label{sec:efficiency}
\label{sec:divergence}

Recall that users must mint $A$- and $B$-tranches in equal proportions. Consequently, they will only be able to reach token allocations in line with their risk preferences if there is a liquid market. Insufficient liquidity would lead to large price spreads (or slippage). Hence, there is a need for market makers, or more generally liquidity providers.

In what follows, we analyze the incentives for liquidity provision of $A$- and $B$-tranches on constant product market makers (CPMM), a special form of automated market makers (AMM) \cite{Mohan.2022}. Note, that CPMMs are only one of many possibilities; tranche token markets could emerge on any trading infrastructure. However, there are a few reasons why CPMMs are of particular importance. \textit{First}, they usually handle a large part of the on-chain trading volume. \textit{Second}, CPMMs allow for composable calls and will always be able to quote a price for any (input) amount. \textit{Third}, CPMMs can be set up in a completely decentralized way and are therefore in line with the strict decentralization requirement of our insurance protocol. 

In a CPMM setup, profitability for liquidity providers is determined by two opposing effects. On the one hand, the pool accumulates protocol fees. The gains are assigned proportionally to all liquidity provision shares. The rate of return depends on the pool's trading volume relative to the pool's liquidity. On the other hand, liquidity providers are s.t. divergence loss (also known as impermanent loss). Divergence loss refers to the problem that liquidity providers lose value, if the liquidity redemption price ratio differs from the liquidity provision price ratio. Intuitively, this effect can be thought of as negative arbitrage. Divergence loss is zero if the two pool tokens maintain their initial price ratio and increases when the relative price is shifting in one direction. 

To assess the incentives for $A$- and $B$-tranche liquidity providers we have to understand divergence loss in the context of our tranche tokens. 
Let us assume a standard $a\cdot b=k$ setup, where $a$ and $b$ represent the initial amount of $A$ and $B$ tokens in the pool and $k$ is a constant product, that determines all feasible combinations of $a$ and $b$. Let us rearrange the equation and take the partial derivative w.r.t. $a$. The absolute value of the resulting slope can be reinterpreted as the relative price.

\begin{equation}
	p_{_{AB}}=  \frac{k}{{a}^2}
	\label{eq:priceratio}
\end{equation}

Trading activity may shift the token allocation to $a^*$ and $b^*$, with $a^* \cdot b^* = k$. Using \eqref{eq:priceratio} we obtain the new price ratio $p^*_{_{AB}}$. This allows us to express the post-trade quantities as a function of the new price ratio $p^*_{_{AB}}$.

\begin{equation}
	a^* = \sqrt{\frac{k}{p^*_{_{AB}}}} \text{ , } \text{ } \text{ } b^* = \sqrt{k \cdot p^*_{_{AB}}}
	\label{eq:quantities}
\end{equation}
We can now compute portfolio values $V_p$ of a simple buy and hold strategy \eqref{eq:bhvalue} with the outcome of liquidity provision \eqref{eq:lpvalue}. 

\begin{align}
	V_P(a, b) &= p^*_{_{AB}} \cdot a + b
	\label{eq:bhvalue} \\
	V_P(a^*, b^*) &= p^*_{_{AB}} \cdot a^* + b^*
	\label{eq:lpvalue}
\end{align}

Using \eqref{eq:quantities} to substitute quantities in \eqref{eq:bhvalue} and \eqref{eq:lpvalue} we get

\begin{align}
	V_P(a, b) &= p^*_{_{AB}} \cdot \sqrt{\frac{k}{p_{_{AB}}}} + \sqrt{k \cdot p_{_{AB}}}, \label{eq:bhvalueplugged}\\
	V_P(a^*, b^*) &= 2 \cdot \sqrt{k \cdot p^*_{_{AB}}}. \label{eq:lpvalueplugged}
\end{align}
Divergence loss can be expressed as follows 
\begin{align}
		D &:= \left| \frac{V_P(a^*, b^*) - V_P(a, b)}{V_P(a, b)} \right |
		\label{eq:dl}
\end{align}
From  \eqref{eq:dl} we plug in \eqref{eq:bhvalueplugged} and \eqref{eq:lpvalueplugged}. After rearranging we get

\begin{equation}
	D = \left| \frac{2 \cdot \sqrt{\frac{p^*_{_{AB}}}{p_{_{_{AB}}}}} - \frac{p^*_{AB}}{p_{_{AB}}} - 1}{\frac{p^*_{AB}}{p_{_{AB}}} + 1} \right|.
	\label{eq:dlPlugged}
\end{equation}

We can now use this equation to analyze two distinct outcomes and observe the effects on the pool and the liquidity providers. \textit{First}, assume the cover is not needed. The contract enters \texttt{Liquid} state, and $A$- and $B$-tranches can be redeemed for equal amounts of $C$. We refer to this case as the \textit{standard case}.
\textit{Second}, assume one of the underlying yield-generating protocols suffers losses. These losses will be reflected in the price of tranche $B$ and therefore have an effect on the liquidity pools that contain $B$. We refer to this as the \textit{benefit case}.

\begin{figure}[h!]
	\centering
	\begin{tikzpicture}
		\draw [->,thick] (0.5,0.8) -- (8,0.8) node (time) [right] {$t$};
		
		\draw (1,0.4) node [below, align=center] {$S$ \\ \texttt{invest()}} -- (1,3.3) ;
		\draw (7,0.4) node [below, align=center] {$T_{1}$ \\ \texttt{divest()}} -- (7,3.3);
		
		\draw [](1,2) node[left] {\small{$p_{_C}$}} -- (7,2);
		\draw[blue] (1,2.5) node[left,blue] {\small{$p_{_A}$}} -- (7,2.8);
		\draw[red] (1,1.5) node[left,red] {\small{$p_{_B}$}} -- (7,2.8);
		
		\draw [<->,thick] (7.2, 2) -- node[right, align=left] {\small{Interest}} (7.2,2.8) ;
		
	\end{tikzpicture}	
	\caption{Relative price development of $A$ and $B$ shares between $S$ and $T_1$, compared to the price of the underlying redeemable asset $C$.}
		\label{fig:lpprice}
\end{figure}
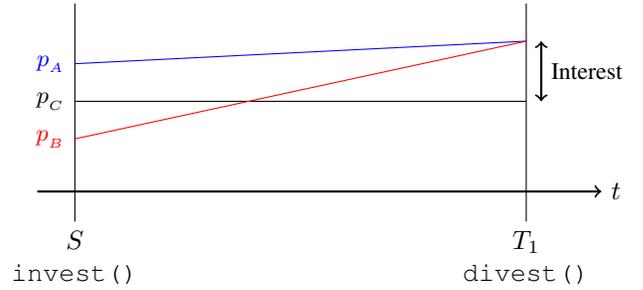

\subsubsection{Standard Case} 
\begin{figure}[h!]
	\center
	\includegraphics[width=8.5cm]{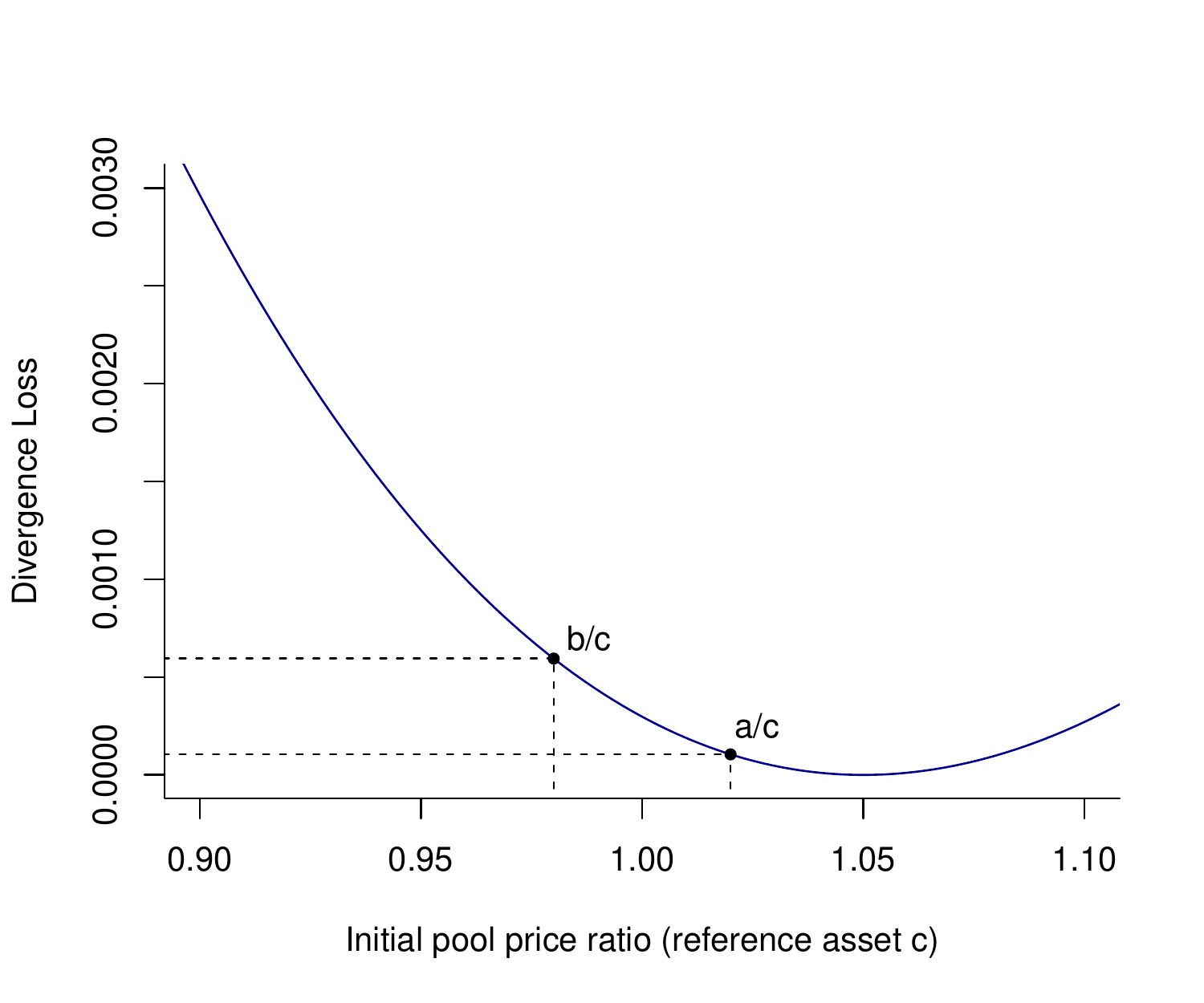}
	\caption{Divergence Loss (in line with equation  \eqref{eq:dlPlugged}) for $a/c$-and $b/c$-Pools with an expected interest of 5\%. The two points marked in our graph represent an example for an initial price spread between $A$ and $B$. The initial valuation of each $a$ token starts at 1.02 $c$, and the valuation of each $b$ token at 0.98 $c$.}
	\label{fig:abcPoolCurve}
\end{figure}

In the standard case $A$-tranches lose their cover value over time. Conversely, $B$-tranches become less risky and will eventually be redeemable for an equal amount of $C$ as $A$-tranches. Hence, we know that $p^*_{_{AB}}=1$. Making use of substitution in \eqref{eq:dlPlugged}, the expected divergence loss can be expressed as a function of the initial price ratio $p_{_{AB}}$. The greater the initial risk premium, the higher the divergence loss for liquidity provision in $A/B$-pools.
Alternatively, a liquidity provider could decide to contribute to an $A/C$- or $B/C$-pool. In $T_1$, we know that $p_A = p_B = p_C \cdot (1+i)$, where $i$ is the accumulated interest. Hence, we know that $p^*_{_{AC}}= p^*_{_{BC}}=1+i$. If we plug this value into \eqref{eq:dlPlugged}, the expected divergence loss, for any expected interest rate, can be expressed as a function of the initial price ratio $p_{_{AB}}$. 
Figure \ref{fig:lpprice} shows the price relations of the three tokens. For $A/B$-pool liquidity provision considerations, interest rates can be neglected. However, for $A/C$- and $B/C$-pools, interest plays an important role. 
Note that $B$-tranche prices already have a positive time trend. As such, interest will further increase the price spread to $C$. Conversely, $A$-tranche prices have a negative time trend and interest will therefore decrease the spread. Consequently, any (positive) interest will create a situation where the divergence loss of $B/C$-pools is greater than the divergence loss of $A/C$-pools. This is shown in Figure \ref{fig:abcPoolCurve}.

While the extent of the divergence loss depends on various factors, it is important to understand that the effect is relatively small. Moreover, there are ways to mitigate a trend-based divergence loss. Alternative pool models, such as the \textit{constant power sum invariant}\cite{niemerg2020yieldspace} can be used to design decentralized exchanges that are better suited for tokens with an inherent price trend.

\subsubsection{Benefit Case}
If any of the yield-generating protocols suffer a loss, $A$-tranche holders will be compensated at the expense of $B$-tranche holders. In extreme scenarios, where one of the yield-generating protocols loses its entire collateral, $B$-tranches become worthless. From \eqref{eq:dlPlugged} we know that $\lim_{p^*_{{AB}} \to \infty } D = -1 $. Hence, $A/B$- and $B/C$-pool liquidity providers are at risk of losing their entire stake.
While this constitutes an additional risk for providers of $B$-tranche liquidity, where they have to expose the $B$ counterpart to an additional risk and effectively stake twice the amount, they receive trading fees in return. As such, the incentives depend on the specifics and the risks of the insured protocols as well as the relative trading volume. 
In extreme cases, where $A/B$ and $B/C$ liquidity provision would be prohibitively risky, liquidity providers could instead contribute to $A/C$-pools. Liquid $A/C$-pools would be sufficient, in the sense that anyone who is interested in coverage could obtain it directly from the pool. This scenario will be further discussed in Section \ref{sec:discussion}.

\section{Discussion}
\label{sec:discussion}

In the introduction we argued that current smart contract-based insurance protocols face various challenges and limitations. We will start our discussion by revisiting these points and explain how our model addresses them. 

First, the vast majority of existing insurance protocols allows for {over-insurance}, where users can buy cover that exceeds their exposure. This can create problematic incentives and -- depending on the jurisdiction -- result in conflict with the law. 
Our model does not allow for over-insurance. The risk and capital are linked through our tranches and cannot be separated without the use of another protocol.

Second, there are various challenges relating to claim assessment. 
All of the existing insurance protocols we have examined have some form of dependency on external factors during the claim assessment process. These dependencies can be introduced through parametric triggers, oracles, community voting or decisions by a predetermined expert council. All of these approaches can lead to undesirable outcomes. The incentives may not be aligned and create situations that can result in deviations from the true outcome. 
In our model, we do not rely on claim assessors, voting in a decentralized autonomous organization (DAO), expert councils, oracles or any trigger events. Instead, we use a deterministic distribution schedule of a common underlying (Liquid Mode) and a sequential choice model in accordance with the seniority of the tranches (Fallback Mode). Consequently, payouts are not conditional on any subjective decisions by an involved- or third-party.

Third, we argued that many DeFi insurance protocols suffer from capital inefficiencies and there certainly is a trade-off between {capital efficiency}, security and special privileges. We found that most existing protocols tend to be conservative or cautious in their approach. The collateral is usually held in low-risk, non-interest-bearing assets. As a result, these protocols have at most 50\% capital efficiency before leverage. 
Some protocols are capable of increasing their efficiency by covering multiple -- ideally uncorrelated -- risks with the same collateral; however, they still require the collateral to be in a low-risk, non-interest-bearing asset. In our model it is possible to hold the collateral, i.e., the $B$-tranche, in a interest-bearing asset without any significant drawbacks on the security side, if the risks of the insured protocols are indeed uncorrelated. Moreover, our approach is quite flexible in the sense that further leverage, based on a larger number of underlying protocols is feasible and could be implemented as an extension.

In addition to these three initial points, there is another advantage related to the risk premium that we came across in the course of our research.
As shown in Section \ref{sec:divergence}, both our cover and collateral ($A$- and $B$-tranches) are freely tradable. The risk premium is simply determined by the relative price between the two tranches. This allows us to create a market-based price-finding mechanism for a fair risk premium. The price can emerge naturally and does not depend on preset parameters or statically implemented risk spreads that may paralyze risk transfer activity.

In Section \ref{sec:divergence} we show that there are greater incentives to provide liquidity for the $A$-tranches than for the $B$-tranches. Even in an extreme case, where the $B$ liquidity would be very low to non-existent, one could still obtain $B$-tranches. To do so, they call the \texttt{splitRisk()} function to mint $A$- and $B$-tranches in equal amounts and then sell the $A$-tranches, for which the market can be assumed to be sufficiently liquid. Anyone interested in the insurance cover could simply buy $A$-tranches on the open market and would not have to interact with the protocol. Assuming a constant supply, greater demand for $A$-tranches would increase the risk spread and therefore incentivize the creation of additional $A$- (and $B$-) tranches.

There are many benefits to our proposal and we believe that this paper significantly contributes towards the DeFi protocol stack. However, every proposal also has its limitations and drawbacks. In the remainder of this section, we discuss some of these limitations and propose potential extensions and new research avenues to mitigate these issues. 
 
First, our model requires a {common underlying} among all involved protocols. The reason for this is to eliminate any reliance on external price sources, i.e., oracles. In liquid mode, we redeem everything to denominations of a unified underlying at the end of a predefined time period. While it is theoretically possible to wrap tokens to give them an arbitrary underlying, this will have one of two consequences: either a dependency on external price sources has to be introduced, or the fallback case in our model would introduce an insurance against relative price movements of the assets and the underlying. The latter may be desirable in some cases, but it is not the default behaviour we want to achieve.

Second, our protocol has a fixed time span. Consequently, insuring assets over a longer period of time requires regular actions from all involved parties. A new contract has to be deployed for each period and the assets need to be moved over. This problem is exacerbated by shorter insurance periods. Longer insurance periods on the other hand increase the time that claimants have to wait for their compensation in case of an incident and also increase the risk of both protocols failing during the same period. We believe this limitation could be mitigated with an extension to the protocol, which uses short insurance periods and rolls over any non-redeemed tranches to a new insurance period. However, an extension of this nature could significantly increase the complexity of the protocol and would require further research to determine the practicality and potential consequences.

Third, in our model we specify minting and redeeming time windows for the tranches. Consequently, the total supply of $A$- and $B$-tranches cannot change during the main insurance period. This can be an issue, especially if there is insufficient liquidity for the $B$-tranches, as discussed in Section \ref{sec:divergence} or if the demand for cover changes significantly. Further research into this topic is necessary, but we believe that under certain circumstances, the minting window could be extended to allow the creation of new shares during the active insurance phase. One requirement for this would be a way to track the accrued interest on the insurance protocol and to increase the costs of the newly created tranches accordingly. Similar considerations can be made for the redeeming window. Early redemption of equal parts of $A$- and $B$-tranches should be possible without large changes to the model. Even early redemption of just $A$-tranches is theoretically possible.

Finally, our model and the reference implementation use two protocols. This is not a strict limitation. In fact, it can be shown that the model works as described as long as the number of tranches is equal to the number of insured protocols. For example, an extension to three protocols is possible with the introduction of a third tranche, without any fundamental changes to the protocol.

A more challenging extension is the addition of further protocols without any changes to the number of tranches. This extension would severely increase the complexity of fallback mode. Recall that $A$-tranche holders get to choose which of the remaining interest-bearing tokens they want to redeem. In a world where the number of tranches is equal to the number of protocols, this is unproblematic, since there will always be sufficient collateral of any type for $A$-tranche holders to choose from. In a model where the number of protocols is greater than the number of tranches, $A$-tranche holders might compete with each other and race to redeem the more valuable collateral. 
As such, models where the number of protocols is greater than the number of tranches can create a first mover advantage, where $A$-tranche holders are treated inconsistently. A potential solution to solve this issue is a two-step approach, that lets tranche holders choose and commit their redemption preferences before the final redemption ratios are calculated.

\section{Conclusion}
\label{sec:conclusion}
In this paper, we propose a fully decentralized DeFi insurance model that does not rely on any external information sources, such as price feeds (oracles) or claim assessors. The general idea of our insurance protocol is to pool assets from two third-party protocols, and allow users to split the pool redemption rights into two freely tradable tranche tokens: $A$ and $B$. Any losses are first absorbed by the $B$-tranche holders. 
$A$-tranche holders will only be negatively affected if 50\% or more of the pooled funds are irrecoverable, or if both protocols become temporarily illiquid and face (partial) losses.

The market for $A$- and $B$-tranches determines the fair risk premium for the insurance.

Our approach has several advantages over other DeFi insurance solutions. In addition to being fully decentralized and trustless, it also prevents over-insurance, does not rely on any parametric triggers, and is highly capital-efficient.

We provide a complete reference implementation of the insurance protocol in Solidity, with coverage for two popular lending market protocols.

We believe that fully decentralized and trustless infrastructure is crucial and may create more transparent, open and resilient financial markets. Our contribution should be seen as a composable building block and a foundation for further research and development efforts.


\section*{Acknowledgment}
The authors would like to thank Tobias Bitterli, Mitchell Goldberg, Emma Littlejohn, Katrin Schuler and Dario Thürkauf.

\appendix
The full Solidity source code for our reference implementation can be found in our github repository: \url{https://github.com/cifunibas/decentralized-insurance}

\vspace{12pt}

\bibliographystyle{IEEEtran}
\bibliography{DeFi_Insurance.bib}

\vspace{12pt}

\end{document}